\documentclass[lettersize,journal]{IEEEtran}
\IEEEoverridecommandlockouts

\usepackage{cite}
\usepackage{algorithmic}
\usepackage[linesnumbered,ruled,vlined]{algorithm2e}
\algsetup{linenosize=\small}
\usepackage{graphicx}
\usepackage{multirow}
\usepackage{textcomp}
\usepackage{xcolor}
\usepackage{scalerel}
\usepackage{graphics}
\usepackage[utf8]{inputenc}
\usepackage{amsfonts}
\usepackage[fleqn]{amsmath}
\usepackage{tablefootnote}
\usepackage{threeparttable}
\usepackage[hidelinks]{hyperref}
\usepackage{array}
\usepackage{multirow}
\usepackage{booktabs}
\usepackage{color}
\usepackage{hhline}
\usepackage{soul}
\usepackage{comment}
\usepackage{tablefootnote}
\usepackage{url}
\usepackage{stfloats}
\usepackage{adjustbox}

\usepackage{tikz}
\usetikzlibrary{arrows,shapes,positioning,decorations.markings}
\usepackage{circuitikz}
\usetikzlibrary{spy,calc}
\usepackage{tikz-timing}
\usetikztiminglibrary{arrows,either,overlays,clockarrows,columntype}

\usetikzlibrary{svg.path}

\definecolor{orcidlogocol}{HTML}{A6CE39}
\tikzset{
    orcidlogo/.pic={
        \fill[orcidlogocol] svg{M256,128c0,70.7-57.3,128-128,128C57.3,256,0,198.7,0,128C0,57.3,57.3,0,128,0C198.7,0,256,57.3,256,128z};
        \fill[white] svg{M86.3,186.2H70.9V79.1h15.4v48.4V186.2z}
        svg{M108.9,79.1h41.6c39.6,0,57,28.3,57,53.6c0,27.5-21.5,53.6-56.8,53.6h-41.8V79.1z M124.3,172.4h24.5c34.9,0,42.9-26.5,42.9-39.7c0-21.5-13.7-39.7-43.7-39.7h-23.7V172.4z}
        svg{M88.7,56.8c0,5.5-4.5,10.1-10.1,10.1c-5.6,0-10.1-4.6-10.1-10.1c0-5.6,4.5-10.1,10.1-10.1C84.2,46.7,88.7,51.3,88.7,56.8z};
    }
}

\newcommand\orcidicon[1]{\href{https://orcid.org/#1}{\mbox{\scalerel*{
                \begin{tikzpicture}[yscale=-1,transform shape]
                \pic{orcidlogo};
                \end{tikzpicture}
            }{|}}}}

\def\BibTeX{{\rm B\kern-.05em{\sc i\kern-.025em b}\kern-.08em
    T\kern-.1667em\lower.7ex\hbox{E}\kern-.125emX}}

\begin{document}
\bstctlcite{IEEEexample:BSTcontrol}
\title{SCALLER: \textbf{S}tandard \textbf{C}ell \textbf{A}ssembled and \textbf{L}ocal \textbf{L}ayout \textbf{E}ffect-based \textbf{R}ing Oscillators}

\author{Muayad~J.~Aljafar\textsuperscript{\orcidicon{0000-0001-7336-4444}}, \IEEEmembership{Member,~IEEE,}
        Zain~Ul~Abideen\textsuperscript{\orcidicon{0000-0002-7065-4400}}, \IEEEmembership{Graduate Student Member,~IEEE,}\\
       Adriaan~Peetermans\textsuperscript{\orcidicon{0000-0002-1211-8888}}, \IEEEmembership{Graduate Student Member,~IEEE,} Benedikt~Gierlichs\textsuperscript{\orcidicon{0000-0002-5866-1990}},
       Samuel~Pagliarini\textsuperscript{\orcidicon{0000-0002-5294-0606}}, \IEEEmembership{Member,~IEEE}
  \thanks{Manuscript received 04 February 2024; revised 13 April 2024; accepted 01 June 2024. This work has been supported by European Union’s H2020 research and innovation programme under grant agreement No 952252 (SAFEST). 
(Corresponding author: S. Pagliarini)}
   \thanks{Samuel Pagliarini is with the ECE Department, Carnegie Mellon University, Pittsburgh, PA (e-mail: pagliarini@cmu.edu).}
   \thanks{Samuel Pagliarini, Muayad J. Aljafar, and Zain Ul Abideen are with the Department of Computer Systems, Centre for Hardware Security, Tallinn University of Technology (TalTech), 12616, Tallinn, Estonia (e-mail: samuel.pagliarini@taltech.ee; muayad.al-jafar@taltech.ee; zain.abideen@taltech.ee).}

  \thanks{Adriaan Peetermans and Benedikt Gierlichs are with imec-COSIC, KU Leuven, Kasteelpark Arenberg 10, Leuven Heverlee, Belgium, B-3001  (e-mail: adriaan.peetermans@esat.kuleuven.be; benedikt.gierlichs@esat.kuleuven.be).}
}

\markboth{SCALLER: Standard Cell Assembled and Local Layout Effect-based Ring Oscillators}{M. Aljafar \MakeLowercase{\textit{et al.}}: SCALLER: \textbf{S}tandard \textbf{C}ell \textbf{A}ssembled and \textbf{L}ocal \textbf{L}ayout \textbf{E}ffect-based \textbf{R}ing Oscillators}

\maketitle

\begin{abstract}
This letter presents a technique that enables very fine tunability of the frequency of Ring Oscillators (ROs). Multiple ROs with different numbers of tunable elements were designed and fabricated in a 65nm CMOS technology. A tunable element consists of two inverters under different local layout effects (LLEs) and a multiplexer. LLEs impact the transient response of inverters deterministically and allow to establish a fine tunable mechanism even in the presence of large process variation. The entire RO is digital and its layout is standard-cell compatible. We demonstrate the tunability of multi-stage ROs with post-silicon measurements of oscillation frequencies in the range of 80-900MHz and tuning steps of 90KHz.

\end{abstract}


\section{Introduction}

In modern System-on-Chip (SoC) designs, various blocks with specific synchrony and frequency requirements are commonly found. Complex SoCs exhibit multiple clock domains, high-frequency demands in the GHz range, and also display interactions between those domains. Therefore, ideal clock generators should be on-chip, providing reliable clock references, and optimized for area and power efficiency.

Ring Oscillators (ROs) are commonly used in Phase-Locked Loop (PLL) designs due to their compact size \cite{compact}. However, even if ROs are identically designed, they are susceptible to significant manufacturing process variation. PLL designs achieve clock robustness through various coarse-grain tuning methods that employ current, voltage, or capacitive load tuning. Such tuning methods typically require custom design. ROs are also utilized in True Random Number Generator (TRNG) designs, where precise control over the RO frequency is crucial for achieving good timing resolution as well as adequate jitter and throughput \cite{coso, tero}.

This letter address the problem of fine-grain tunability in ROs, which benefits both PLL designs and TRNG design. Recently, there has been an increasing adoption of digital architectures in all-digital PLLs \cite{compact, 6177036, 6757428}, which are still susceptible to frequency fluctuations due to process variation and therefore require \textbf{digital tunability} capability. For some TRNG architectures, precise RO frequency tuning is essential to mitigate the effects of process variation and deliver random numbers that are not biased. 

This work introduces a novel technique for tuning the frequency of ROs using Local Layout Effects (LLEs). Traditionally, LLEs are considered detractors as they can affect a transistor's performance based on its surroundings. However, we systematically control and leverage these effects, particularly the Well Proximity Effect (WPE), in order to achieve enhanced control over the RO frequency. The proposed technique utilizes only regular transistors arranged as standard cells, providing a flexible alternative to custom designs.

\section{Background and Proposed Approach}
\label {Methodology}

\subsection{Well Proximity Effect (WPE)}

\begin{figure}[t]
\centerline{\includegraphics[width=0.6\linewidth]{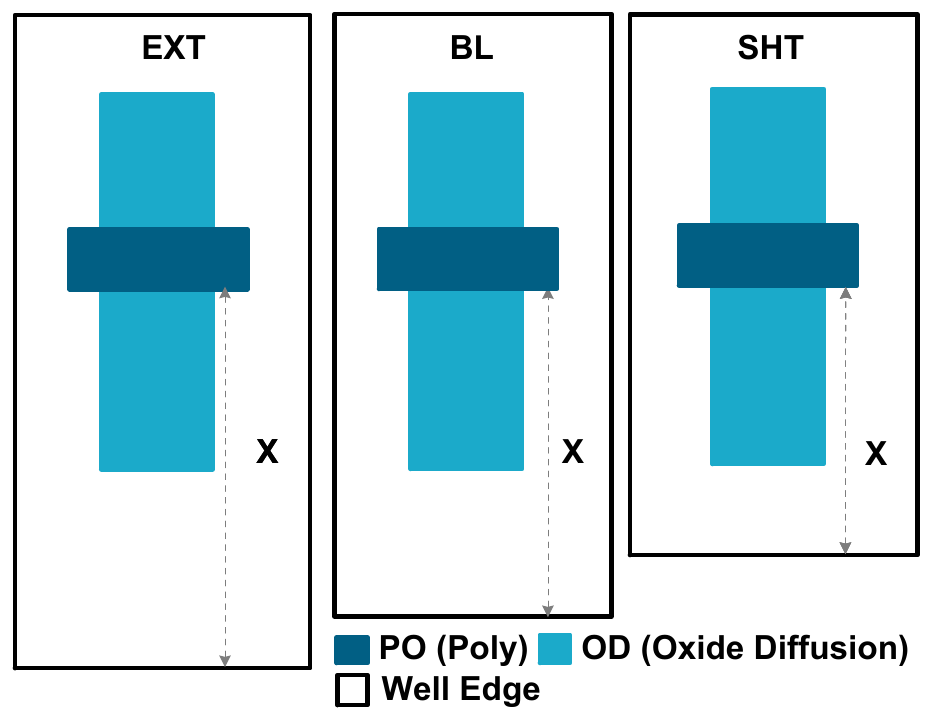}}
\caption{Well Proximity Effect. Three simplified transistor layouts with different values of X are shown.} 
\label{WPE}
\end{figure}

LLEs are a consequence of the reduced process geometries in lithography. Some foundries refer to these as layout-dependent effects or LDEs. WPE is an example of an LLE that appears in sub-100nm CMOS technologies, including FinFET ones. WPE relates to the proximity of a device (transistor) to the well edge. A transistor that is close to the edge will show different performance (larger voltage threshold (V\textsubscript{t}) and smaller drain current) from that of a device located far from the edge (X in Fig.~\ref{WPE}), even if the transistors are drawn with identical dimensions \cite{1224497, 1568798}. This effect is present regardless of the transistor voltage threshold (i.e., standard, high, or low V\textsubscript{t}). Commercial PDKs include transistor electrical models that take LLEs into account. 

\begin{figure}[tbp]
\centerline{\includegraphics[width=0.9\linewidth]{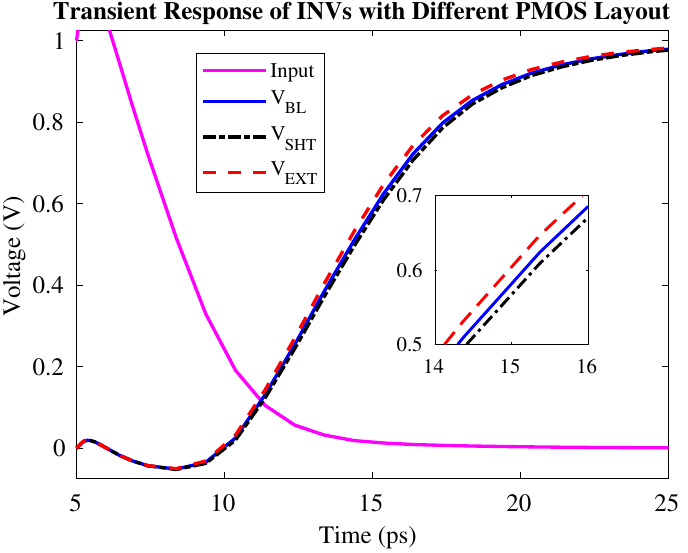}}
\caption{Transient response of three inverters with baseline and manipulated PMOS transistors. Obtained from electrical simulation in Cadence Spectre.}
\label{timeResponse}
\end{figure}

Such effects have been leveraged before in the context of analog circuit obfuscation \cite{aljafar2022leveraging}, i.e., to protect analog Intellectual Property (IP). However, as a general rule, LLEs are considered detractors in conventional circuit design. WPE, in particular, impacts the time response of a transistor \cite{4430259}. As shown in Fig.~\ref{WPE}, three simplified transistor layouts with different well proximity X are drawn, where the baseline (BL) is the unmodified layout from a commercial standard cell library. The well size in BL was manipulated and two variants were created: a shortened (SHT) and an extended (EXT) variant. Note that these are trivial modifications to existing cells that are easily scripted; only one layout layer is modified -- the well layer -- while all others remain untouched. Fig.~\ref{timeResponse} shows the transient response of the baseline versus the two modified variants. Notice that $V_{EXT}$ is faster than $V_{BL}$. In turn, $V_{BL}$ is faster than $V_{SHT}$. We clarify that the three inverters are presented with the same input and have the same load. Hence, the difference in performance assuredly comes from the WPE. Our simulation shows that $V_{BL}$ is 2.86\% faster than $V_{SHT}$ and 2.02\% slower than $V_{EXT}$ when the input/output voltages are at 50\% of VDD. Without loss of generality, we have deliberately focused only on PMOS transistors in our ROs that will be presented next. NMOS transistors can also be targeted in dual- or triple-well technologies.

\subsection{Tunable ROs}

\begin{figure}[tbp]
\centerline{\includegraphics[scale=.56]{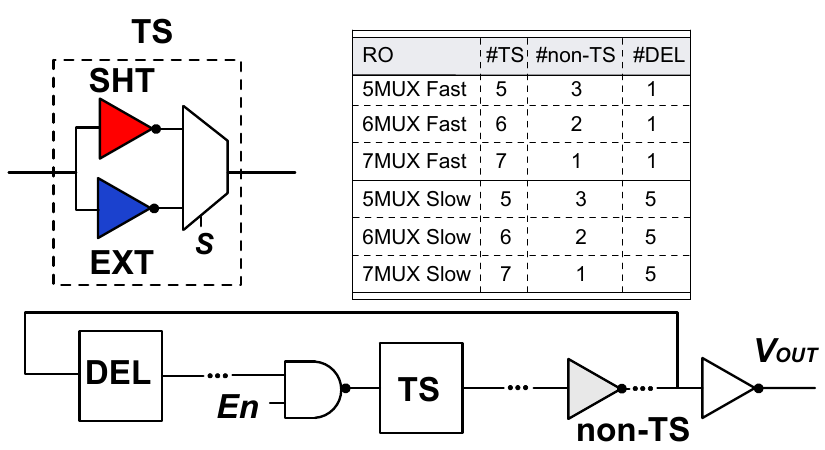}}
\caption{Structure of the proposed tunable stage (TS). Multiple configurations for building an RO were considered. All ROs have 9 stages, 8 from inverters and one from a NAND gate.}
\label{5MUXRO}
\end{figure}

We built 9-stage tunable ROs by a combination of Tunable Stages (TSs), non-Tunable Stages (non-TSs), and delay (DEL) cells. Non-TSs are built from regular inverters that have no well manipulation. DEL cells are starved buffers that slow down the frequency of oscillation to save power (such cells are present in every standard cell library). A TS consists of a 2-input multiplexer (MUX) and inverters with both shortened and extended PMOS wells (SHT and EXT), as shown in Fig.~\ref{5MUXRO}. TSs allow tunability by selecting inverters with different \emph{speeds} in an RO. 

Fig.~\ref{5MUXRO} shows different configurations for building 9-stage LLE-based ROs only from regular standard cells. The ROs may have 5, 6, or 7 TSs, and 1, 2, or 3 non-TSs. Our proposed ROs also come in a Fast and a Slow variant: Slow ROs come with 5 DEL cells while Fast ones come with 1 DEL cell. A NAND gate is always used in every RO (for the enable condition) while also acting as a non-TS. 
To decouple the WPE impact on the RO frequency, a `reference RO' (Ref RO) was also considered. The Ref RO is identical to the LLE RO, apart from the SHT and EXT PMOS wells which are both replaced by BL wells. Ref ROs and LLE ROs are then arranged in pairs. All 6 RO configurations listed in Fig.~\ref{5MUXRO} were designed and validated in a 65nm commercial CMOS technology.

Fig.~\ref{TRNG_5MUXs_layout} shows the layout of a 5MUX Fast pair where the top RO is the Ref RO and the bottom one is the LLE-based RO. The image includes metal lines and the well size of PMOS transistors (normalized). We observe adjustments to the well sizes of SHT PMOS and EXT PMOS, with SHT PMOS reduced by 28\% and EXT PMOS increased by 37\%. Expanding the well size of PMOS transistors in the inverters does not lead to an overall increase in the RO's area, as the `width' of each RO stage is determined by MUXes. This strategy is beneficial since it simplifies the realization of a symmetrical layout. Table~\ref{macro_specs} shows the pre-silicon specs of individual ROs. Notice that the height of all ROs is the same since they take 6 standard cell rows. 

\begin{table}[t]
 	\caption{Pre-silicon Specs of Individual ROs} 
	\begin{center}
		\begin{tabular}{c c c c}
			\hline
			\hline
            {\sc RO}                &   {\sc Area} ($\mu m^2$) &  \multicolumn{2}{c}{\sc Frequency (MHz)}     \\ 
            \cline{3-4}             &                         &    {\sc ref ro}    &  {\sc lle ro}      \\      \hline 
	      {\sc 5mux fast}         &   $24.9 \times 10.8$    &    {\sc 720.3}    &  {\sc 718.5}       \\
	      {\sc 6mux fast}         &   $26.1 \times 10.8$    &    {\sc 668.7}    &  {\sc 666.0}         \\
            {\sc 7mux fast}         &   $27.8 \times 10.8$    &    {\sc 624.5}    &  {\sc 621.1}       \\
            \hline
            {\sc 5mux slow}         &   $51.8 \times 10.8$    &    {\sc 65.29}    &  {\sc 65.24}       \\
	      {\sc 6mux slow}         &   $54.7 \times 10.8$    &    {\sc 64.84}    &  {\sc 64.81}       \\
            {\sc 7mux slow}         &   $56.8 \times 10.8$    &    {\sc 64.41}    &  {\sc 64.39}       \\
			\hline
			\hline						
		\end{tabular}
	\label{macro_specs}
	\end{center}
\end{table}

\section{Chip Design and Measurements} \label {Measurements}

\begin{figure}[t]
\centerline{\includegraphics[width=0.85\linewidth]{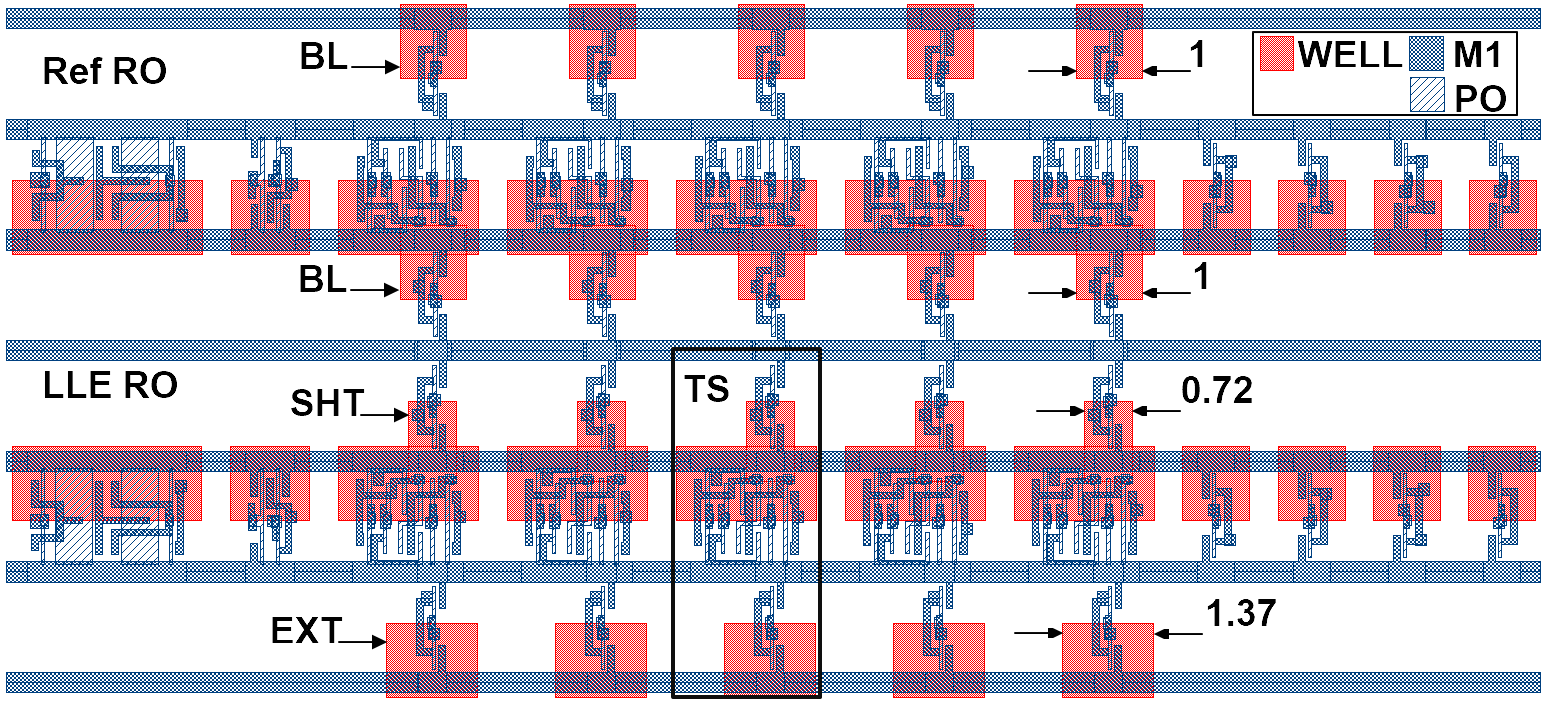}}
\caption{Layout of 5MUX RO pair ($24.9 \mu m\times 10.8\mu m)$ that is compatible with std. cell row-based design. The highlighted TS consists of a MUX (middle) with two manipulated inverters (above and below). Well sizes are normalized.}
\label{TRNG_5MUXs_layout}
\end{figure}

To observe the impact of LLEs on oscillation frequency, we consistently analyze LLE and Ref ROs in pairs. This ensures that both ROs are physically located near each other, thus decreasing the effect of (global) process variation. We refer to a pair of ROs as either `$k$MUX Fast pair' or `$k$MUX Slow pair,' where $k$ denotes the number of MUXes (or TSs) in each RO. For example, a 5MUX Fast pair consists of a pair of ROs with 5 TSs and 1 DEL cell each, while a 5MUX Slow pair consists of a pair of ROs with 5 TSs and 5 DEL cells each. The layout of a pair of ROs can be seen in Figure~\ref{TRNG_5MUXs_layout}, with some portions redacted to protect the foundry IP.

\begin{figure}[tbp]
\centerline{\includegraphics[scale=.26]{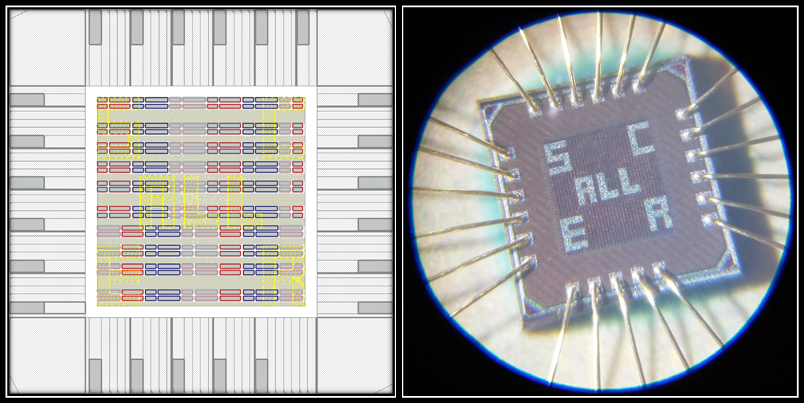}}
\caption{Layout ($960\mu m \times 960\mu m$) and die micrograph. Colorful rectangles in the layout view are the pairs of ROs.}
\label{LayoutChip}
\vspace{-10pt}
\end{figure}

We then conceived a floorplan with a large number of pairs of ROs and fabricated a total of $100$ chips (80 bare dies, 20 packaged dies) in a 65nm commercial technology. Each chip contains $224$ pairs of ROs, from which $116$ are of the Slow type and the rest are Fast ones. The Slow pairs include $40$ 5MUX, $40$ 6MUX, and $36$ 7MUX variants, and the Fast pairs include $36$ 5MUX, $40$ 6MUX, and $32$ 7MUX variants. One pair can be accessed at a time via a shared interface. The oscillation for Ref and LLE ROs is measured directly on the chip IO pins. It is worth noting that no frequency interference was observed, meaning that no frequency locking was detected between the RO pairs. This is an important property for TRNGs where multiple ROs may operate in close vicinity. In either case, for all the measurements herein reported, only one RO of the pair is active at a time. Fig.~\ref{LayoutChip} shows the layout and die micrograph of the chip. We found that WPEs can be leveraged for fine tunability in ROs despite process variation. We also found that LLE and Ref ROs respond to changes in V and T in the same manner; therefore we omit this result. We tested all $20$ packaged chips and found that our measurements are generally in line with our simulation predictions.

\subsection{Power Analysis}
\begin{figure}[tbp]
\centerline{\includegraphics[width=0.95\linewidth]{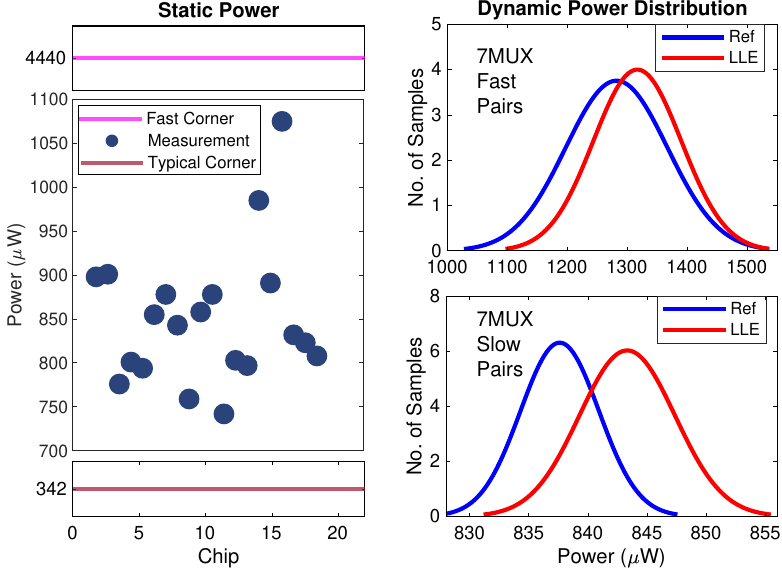}}
\caption{Power analysis. Dots correspond to leakage measurements from all 20 packaged chips. Horizontal lines show power estimation for FF and TT corners from chip-level electrical simulation. Distributions correspond to dynamic power in all 7MUX Fast and Slow pairs of chip \#12.}
\label{PWR}
\vspace{-10pt}
\end{figure}

Figure~\ref{PWR} presents the leakage power measurements for all chips, revealing a significant range of power dissipation from 745${\mu}W$ to 1072${\mu}W$, with a mean of 849${\mu}W$ and a standard deviation of 77${\mu}W$. This wide range suggests that our chips fall somewhere between the TT and FF corners in terms of process variation. Additionally, we conducted dynamic power measurements for all 7MUX ROs of chip \#12, assessing one RO at a time. Fig.~\ref{PWR} also displays the power distributions of 7MUX Fast and Slow pairs, with similar distributions observed for other 6MUX and 5MUX ROs and therefore omitted. Additional experiments where supply voltage was changed revealed no significant difference in the frequency (both LLE and Ref ROs scale with VDD as expected). 

\subsection{Impact of WPE on Tunability}
\begin{figure*}[htb]
\centering
\includegraphics[width=1.0\linewidth]{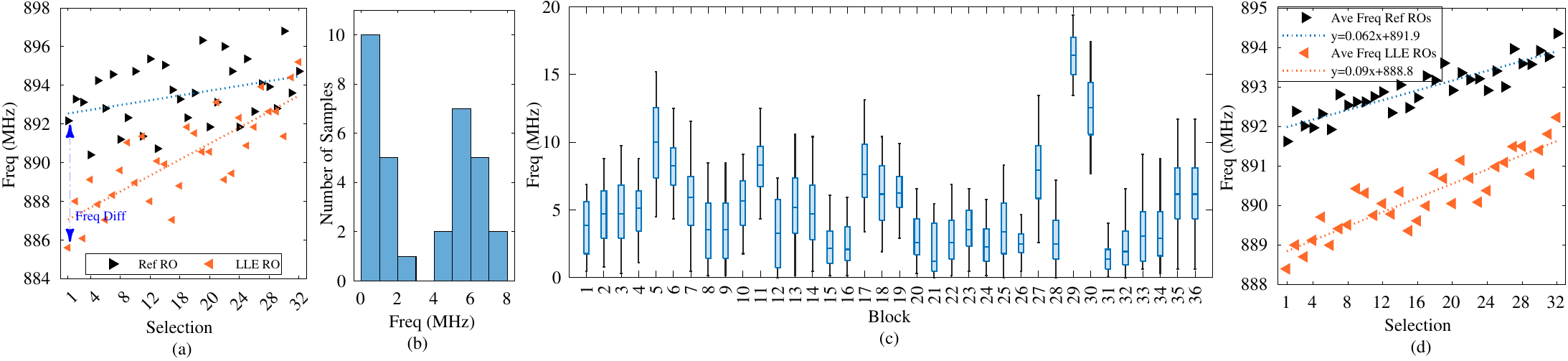}
\caption{Impact of WPE on tunability. (a) The behavior of a 5MUX Fast pair. (b) Distribution of the frequency difference in a 5MUX Fast block. (c) Variations in mean frequency difference in all Fast blocks with 5MUX ROs in chip \#2. (d) Mean frequencies in Fast blocks with 5MUX ROs in the tested chip.}
\label{isolation_5mux}
\end{figure*}

Now, let us analyze RO pairs independently and assess the impact of WPEs on RO tunability by calculating the frequency difference for each possible selection. In Fig.~\ref{isolation_5mux}a, we observe the behavior of a 5MUX Fast RO pair from chip \#2. The scatter plot shows frequencies for every selection of inverters (32 possible selections with 5 MUXes). Data points are arranged based on the number of EXT inverters used in the LLE RO, ranging from none to five. 

Additionally, we calculated the frequency differences in the ROs for each selection, creating a distribution shown in Fig.~\ref{isolation_5mux}b that reveals a tunability range of 8MHz in LLE ROs. This distribution is not normal, which indicates the LLE manipulation is effective. It is important to note that we ensured equal selections in both ROs while evaluating tunability. The box chart in Fig.~\ref{isolation_5mux}c summarizes the distribution of every Fast 5MUX RO (`Block') within a single chip. It shows variations in the tuning range of each block, ranging from 3.7 MHz to 11 MHz, which result from process variation. Fig.~\ref{isolation_5mux}d displays the mean frequencies for every selection in all Fast 5MUX ROs. Each data point represents the average frequency of 36 5MUX Fast ROs on the same chip, arranged in the same manner as shown in Fig.~\ref{isolation_5mux}a. Note that the increase in frequency  per configuration bit in LLE ROs ($y=0.090x$) is larger than in Ref ROs ($y=0.062x$). This difference in the slope of the curve is attributed to the manipulated inverters' layouts in LLE ROs. 

\begin{figure}[tb]
\centering
\includegraphics[width=1.0\linewidth]{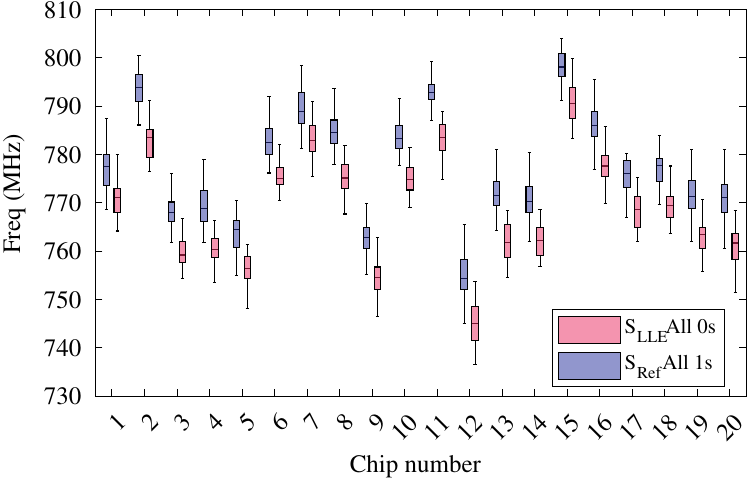}
\caption{Characterization of 7MUX Fast pairs for all the chips.}
\label{fig:char_7mux_fast}
\vspace{-10pt}
\end{figure}

The tunability range in LLE ROs is approximately 3.84 MHz (taken from Fig.~\ref{isolation_5mux}d data by subtracting the rightmost from the leftmost datapoint), with an average tuning step of 90 KHz, indicating a very fine level of tunability. In the Ref RO, the tuning range is 1.92 MHz. Table \ref{tab:results_m_std} presents post-silicon results for 5MUX Fast, 5MUX Slow, 6MUX  Fast, and 7MUX Fast. Each cell in the table displays the mean and standard deviation for the given RO and the given selection bits. Each result corresponds to all data points collected from all ROs (36 5MUX, 40 6MUX, 32 7MUX). Remarkably, the mean frequency for all ROs maintains the same relative order (Ref All 1s $>$ LLE All 1s $>$ Ref All 0s $>$ LLE All 0s), indicating that the tuning mechanism works more favourably to reduce the frequency than to increase the frequency. The data in Table~\ref{tab:results_m_std} reveals that distributions associated with LLE ROs typically exhibit larger standard deviations than those related to Ref ROs.

We have also identified that the 7MUX blocks exhibit a wider range of tunability than others due to having more TSs in their ROs. This range was determined by calculating the mean difference of distributions corresponding to $S_{LLE}$ All 0s and $S_{Ref}$ All 1s, representing selections with only slow transient response inverters in LLE ROs and all control select bits set to 1s in Ref ROs, respectively. This calculation was repeated for 7MUX Fast blocks in all 20 chips, and the results were summarized in the box chart displayed in Fig.~\ref{fig:char_7mux_fast}. In other words, our results are consistent when we look at all 20 chips and consider die to die variation. Table~\ref{tab:results_m_std}, on the other hand, addresses intra-die variation. Both results are consistent.

\begingroup
\setlength{\tabcolsep}{3.0pt} 
\renewcommand{\arraystretch}{1.1} 
\begin{table} [!t]
\scriptsize \centering
\caption{Mean and standard deviation ($\mu$/$\sigma$) for the ROs}
\label{tab:results_m_std}
\begin{tabular}{c|c|c|c|c} \hline
\textbf{RO} & \textbf{Ref All 0s} & \textbf{Ref All 1s} & \textbf{LLE All 0s} & \textbf{LLE All 1s} \\ \hline

5MUX Fast & 897.37/6.09 & 903.37/5.34 & 894.85/5.87 & 902.20/5.90 \\ \hline 
6MUX Fast & 835.60/3.75 & 844.08/3.84 & 833.84/4.15 & 843.46/4.51 \\ \hline 
7MUX Fast & 786.27/3.66 & 793.24/3.96 & 782.83/3.67 & 790.72/4.07 \\ \hline 
\end{tabular}

\end{table}
\endgroup 

\section{Conclusions}
 \label{Conclusions}
We have demonstrated very fine tunability in tunable ROs by leveraging the WPE that tends to bring the frequency difference in RO pairs close to each other. Any of the proposed ROs can be used depending on the range of tunability needed. The measurements confirm the fine tunability of our pairs of ROs whereas the results related to frequency distribution (Tab.~\ref{tab:results_m_std}) indicate the potential of our ROs for use in TRNGs, which is precisely where our future work will focus. For PLL designs, our digital-only tunability approach may prove both adequate and advantageous for all-digital PLLs.

\bibliographystyle{IEEEtran}
\bibliography{tvlsi}

\begin{thebibliography}{1}
\providecommand{\url}[1]{#1}
\csname url@samestyle\endcsname
\providecommand{\newblock}{\relax}
\providecommand{\bibinfo}[2]{#2}
\providecommand{\BIBentrySTDinterwordspacing}{\spaceskip=0pt\relax}
\providecommand{\BIBentryALTinterwordstretchfactor}{4}
\providecommand{\BIBentryALTinterwordspacing}{\spaceskip=\fontdimen2\font plus
\BIBentryALTinterwordstretchfactor\fontdimen3\font minus \fontdimen4\font\relax}
\providecommand{\BIBforeignlanguage}[2]{{%
\expandafter\ifx\csname l@#1\endcsname\relax
\typeout{** WARNING: IEEEtran.bst: No hyphenation pattern has been}%
\typeout{** loaded for the language `#1'. Using the pattern for}%
\typeout{** the default language instead.}%
\else
\language=\csname l@#1\endcsname
\fi
#2}}
\providecommand{\BIBdecl}{\relax}
\BIBdecl

\bibitem{compact}
A.~Musa, W.~Deng, T.~Siriburanon, M.~Miyahara, K.~Okada, and A.~Matsuzawa, ``A compact, low-power and low-jitter dual-loop injection locked pll using all-digital pvt calibration,'' \emph{IEEE Journal of Solid-State Circuits}, vol.~49, no.~1, pp. 50--60, 2014.

\bibitem{coso}
P.~Kohlbrenner and K.~Gaj, ``An embedded true random number generator for fpgas,'' in \emph{12th Symposium on FPGA}.\hskip 1em plus 0.5em minus 0.4em\relax {ACM}, 2004, pp. 71--78.

\bibitem{tero}
M.~Varchola and M.~Drutarovsky, ``New high entropy element for {FPGA} based true random number generators,'' in \emph{CHES}, vol. 6225, 2010, pp. 351--365.

\bibitem{6177036}
P.~Park, J.~Park, H.~Park, and S.~Cho, ``An all-digital clock generator using a fractionally injection-locked oscillator in 65nm cmos,'' in \emph{ISSCC}, 2012, pp. 336--337.

\bibitem{6757428}
W.~Deng, D.~Yang, T.~Ueno, T.~Siriburanon, S.~Kondo, K.~Okada, and A.~Matsuzawa, ``A $0.0066mm^2$ 780uw fully synthesizable pll with a current-output dac and an interpolative phase-coupled oscillator using edge-injection technique,'' in \emph{ISSCC}, 2014, pp. 266--267.

\bibitem{1224497}
T.~Hook, J.~Brown, P.~Cottrell, E.~Adler, D.~Hoyniak, J.~Johnson, and R.~Mann, ``Lateral ion implant straggle and mask proximity effect,'' \emph{IEEE Transactions on Electron Devices}, vol.~50, no.~9, pp. 1946--1951, 2003.

\bibitem{1568798}
Y.-M. Sheu, K.-W. Su, S.-J. Yang, H.-T. Chen, C.-C. Wang, M.-J. Chen, and S.~Liu, ``Modeling well edge proximity effect on highly-scaled mosfets,'' in \emph{Proceedings of the IEEE 2005 Custom Integrated Circuits Conference, 2005.}, 2005, pp. 831--834.

\bibitem{aljafar2022leveraging}
M.~J. Aljafar, F.~Azais, M.-L. Flottes, and S.~Pagliarini, ``Leveraging layout-based effects for locking analog ics,'' in \emph{Proceedings of the 2022 Workshop on Attacks and Solutions in Hardware Security}, 2022, pp. 5--13.

\bibitem{4430259}
T.~Kanamoto, Y.~Ogasahara, K.~Natsume, K.~Yamaguchi, H.~Amishiro, T.~Watanabe, and M.~Hashimoto, ``Impact of well edge proximity effect on timing,'' in \emph{ESSCIRC 2007 - 33rd European Solid-State Circuits Conference}, 2007, pp. 115--118.

\end{thebibliography}

\end{document}